# Propagation Dynamics of Photonic Toroidal Vortices Mediated by Orbital Angular Momenta

Xin Liu[1,2,†], Nianjia Zhang[3], Qian Cao[3], Jinsong Liu[1,2], Chunhao Liang[1,2,‡], Qiwen Zhan[3,‡], Yangjian Cai[1,2,‡]

[1]Shandong Provincial Engineering and Technical Center of Light Manipulations and Shandong Provincial Key Laboratory of Optics and Photonic Device, School of Physics and Electronics, Shandong Normal University, Jinan 250014, China.

[2]Collaborative Innovation Center of Light Manipulations and Applications, Shandong Normal University, Jinan, 250358, China.

[3]School of Optical-Electrical and Computer Engineering, University of Shanghai for Science and Technology, Shanghai 200093, China.

[†]e-mail: cnliuxin1995@gmail.com;

[‡]Corresponding authors: cliang@dal.ca; qwzhan@usst.edu.cn; yangjiancai@sdnu.edu.cn;

**Abstract**:

The dynamics of vortex rings in fluids have long captivated researchers due to the intriguing complexity of their behavior, despite the apparent simplicity of their structure. In optics, photonic toroidal vortices constitute a novel class of three-dimensional, space-time nonseparable structured light fields that carry transverse orbital angular momentum. However, as solutions to the dispersive form of Maxwell's equations, these wavepackets do not survive upon nondispersive propagation, and their dynamics remain elusive. In this article, the dynamics of photonic toroidal vortices under various dispersion regimes, mediated by both transverse and longitudinal orbital angular momentum, are investigated through simulations and experiments. The results reveal that the motion of a toroidal vortex is strongly affected by the presence of longitudinal orbital angular momentum. The swirling flow destabilizes the toroidal structure under dispersion conditions and induces topological transformations in the vortex line characterized by its annihilation and subsequent reformation in vacuum. Remarkably, the renascent toroidal vortex exhibits robust propagation in vacuum while maintaining its toroidal structure. These findings are supported by experimental validation and highlight the potential of photonic toroidal vortices as controllable channels for directional energy and information transfer.

## 1. Introduction

Toroidal vortices, also known as vortex rings, are fascinating three-dimensional (3D) torus-shaped structures characterized by vortical flow confined around a closed vortex line. Such flow patterns are frequently encountered in natural contexts and fluid mechanics [1,2], manifest in various forms, ranging from the turbulent flows of liquids and gases [3-5] such as smoke rings and cavitation bubbles, to thermally induced formations in mushroom clouds [5], and biologically produced vortices associated with microscale locomotion [6], seed dispersal [7,8], and cardiovascular flow [9]. Despite the apparent simplicity of toroidal vortices structure, their evolution exhibits intriguing complexity and has long motivated extensive research aimed at elucidating these dynamics under various flows [10-14].

In the realm of electromagnetics, two types of toroidal light vortices have attracted wide attention since their experimental creations. One is the vector toroidal pulse, which be visualized as a torus made of electric/magnetic field lines [15-17], exhibits nontrivial topological textures [18,19] and propagation-robust dynamics [20]. Another representative structure is the scalar photonic toroidal vortex [21], characterized by a closed-loop spatiotemporal optical vortex (STOV) [22-26]. This configuration gives rise to a $2\pi$ phase variation around the surface of a torus in the poloidal plane, thereby yielding a transverse orbital angular momentum (OAM) density that is tangent to the vortex line, as shown in Fig. 1. A toroidal vortex wavepacket is an approximate solution to the time-dependent wave equation under anomalous dispersion, but it exhibits strongly unstable evolution in free space [21,27]. A toroidal vortex carrying a toroidal swirl forms a scalar optical Hopfion topology [28] which encompasses both transverse and longitudinal OAM. It has been proven that the coupling between these OAMs within a toroidal vortex leads to entirely distinct topological properties in high-harmonic generation [29]. It is amazing that fluid vortex rings are capable of holding their morphology for a long time in a calm medium. However, although the photonic toroidal vortices can now be readily synthesized in free space, their robust free propagation is still out of reach, and their dispersive spreading and motion under the interactions of transverse and longitudinal OAMs are not yet investigated.

Understanding the dynamics and mutual interaction of various types of vortices is a key ingredient in clarifying and taming high-dimensional light. In this study, we reveal that spatiotemporal toroidal vortices exhibit nontrivial dispersive dynamics, mediated by the coupling between their transverse and longitudinal OAM. In the absence of longitudinal OAM, the toroidal vortex evolves similarly to a conventional STOV pulse. However, the inclusion of a longitudinal OAM disturbs the flow of toroidal vortex, fundamentally alters the dispersive behavior, leading to several distinctive phenomena: (1) the toroidal vortex becomes highly unstable in both normally and anomalously dispersive media; (2) in vacuum, the vortex line undergoes breakdown and subsequent reconstruction during propagation, and the renascent toroidal vortex robustly preserves its morphology over extended distances. These findings are confirmed in the experiments.

## 2. Numerical simulation results and analysis

The complex field of a toroidal vortex wavepacket at $z = 0$ can be expressed as

$$\Psi(x, y, \tau) = \left(\frac{\sqrt{2}\rho_\perp}{w_0}\right)^{|\ell_1|} \exp\left(-\frac{\rho_\perp^2}{w_0^2}\right) L_p^{|\ell_1|}\left(\frac{2\rho_\perp^2}{w_0^2}\right) \exp\left[-i\left(\ell_1 \varphi_\perp + \ell_2 \gamma_s\right)\right], \tag{1}$$

where $L_p^{|\ell_1|}(\cdot)$ is the associated Laguerre polynomial. $\rho_\perp = \sqrt{(r_s - r_0)^2 + \tau^2}$, $\varphi_\perp = \tan^{-1}[\tau/(r_s - r_0)]$ are defined in the *poloidal plane* and $r_s = \sqrt{x^2 + y^2}$, $\gamma_s = \tan^{-1}(y/x)$ are defined in the *toroidal plane*; $r_0$ is a constant that quantifies the radius of the toroidal vortex line; $\ell_1$ and $\ell_2$ are the winding numbers that quantify the twist number of helicoid phases at the poloidal and toroidal plane, respectively. Figure 1(a) displays the 3D structure of a photonic toroidal vortex with $\ell_2 = 0$, which has a closed vortex line. Since $\langle L_\tau \rangle \propto \ell_2 = 0$, the OAM is oriented tangentially to the vortex line, resulting in an energy flow pattern that is identical to that of a standard STOV (see Appendices A and B). When $\ell_2 \neq 0$, Eq. (1) describes the wave function of a scalar optical Hopfion, whose phase structure can be visualized as a torus formed by nested equiphase lines, as shown in Fig. 1(b)-(d) for different $\ell_2$. A toroidal vortex with $\ell_2 \neq 0$ could simultaneously carry transverse OAM [$\langle \hat{L}_x \rangle$ and $\langle \hat{L}_y \rangle$ in the poloidal plane] and longitudinal OAM [$\langle \hat{L}_\tau \rangle \propto \ell_2 \neq 0$ in the toroidal plane, see Appendix C]. The poloidal flow (dependent on dispersions) interacts with toroidal swirl, yielding a complex energy flow (Appendix B).

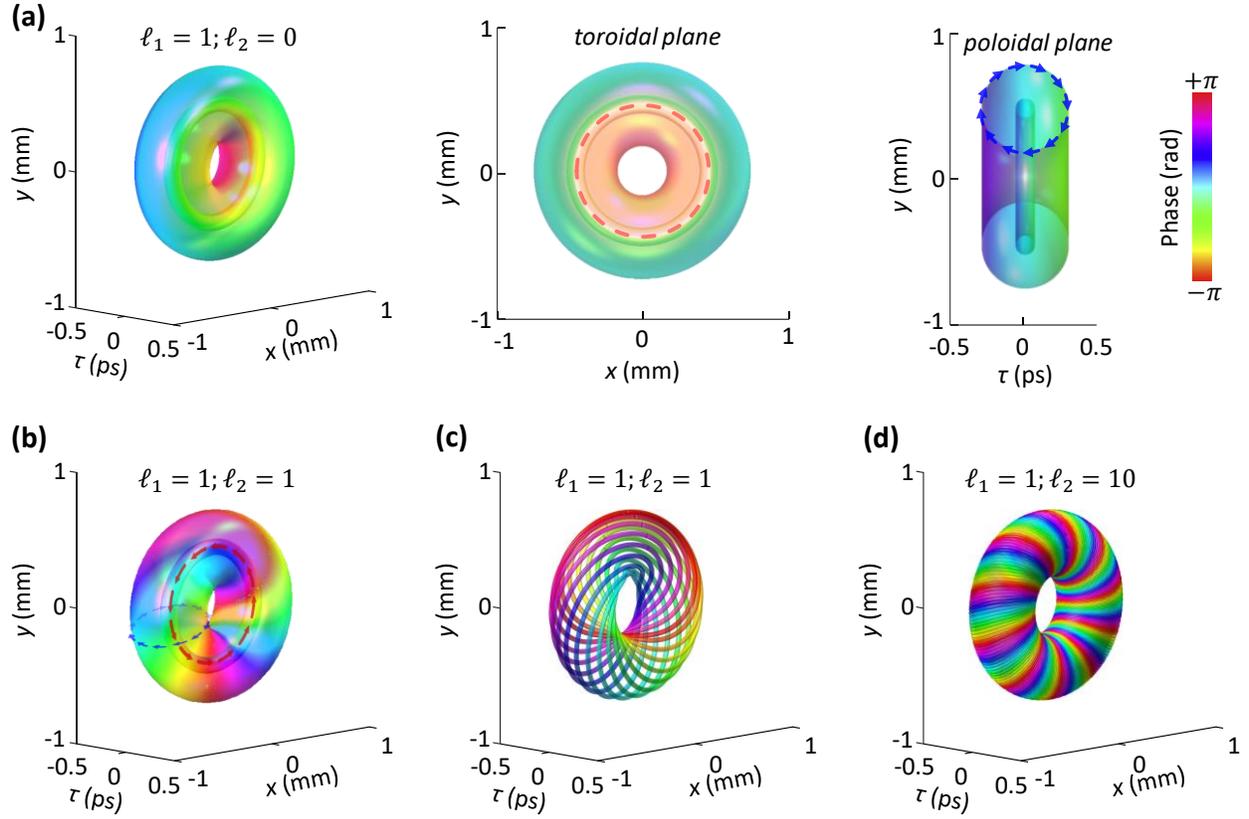

FIG. 1 (a) Iso-intensity profile and phase map of a toroidal vortex of $\ell_1 = 1$ and $\ell_2 = 0$ in different perspective views. (b) Iso-intensity profile and phase map of a photonic toroidal vortex of $\ell_1 = 1$ and $\ell_2 = 1$. (c) and (d) The equiphase representation of a Hopfion wavepacket with different $\ell_2$. The color denotes phase. The parameters are $\lambda_0 = 800$nm $w_\tau = 1$ps, $w_x = 1$mm, $w_0 = 0.25$. $r_0 = 0.5$mm.

In simulations, we use the angular spectrum relation [see Appendix C] to numerically calculate the full space-time wavepacket of toroidal vortices in dispersive media. Figures 2-4(a) show the propagation dynamics of toroidal vortices without a longitudinal OAM under different dispersion cases. In all cases, the ring-shaped structure of the toroidal vortex collapses during propagation as a result of spatial diffraction within the toroidal plane. In contrast, for toroidal vortices carrying longitudinal OAM, the ring-shaped structure is preserved, as shown in Figs. 2–4(b) and 2-4(c). This difference arises from the fact that the longitudinal OAM induces a swirling flow in the toroidal plane, which helps preserve the ring-shaped structure. However, this swirling flow interacts with the poloidal flow, giving rise to more complex spatiotemporal dynamics, as will be discussed in detail below.

In a normally dispersive medium ($\beta_2 = k_0^{-1}$), the vortex line carried by the toroidal vortex without longitudinal OAM in Fig. 2(a) becomes stretched, finally causing it to split and peel off two separated pieces in time. The phase difference between the two pieces approaches $\pi$ near the fissure because of the "saddle" energy flow [see Appendixes A and B]. But in Fig. 2(b), the toroidal vortex with longitudinal OAM initially breaks into two separate pieces over a short propagation distance. After that, the vortex line quickly reconstructs, and the entire wavepacket exhibits a well-defined toroidal structure. The interaction between saddle- and spiral-shaped energy flows leads to a polarity reversal of the reconstructed vortex line. While the structure at this position is highly unstable, as propagation further proceeds in Fig. 2(c), the toroidal vortex undergoes a second breakup, splitting into two independent pieces. In contrast to the evolution in Fig. 2(a), this peeling off occurs in a reversed direction due to the polarity reversal of the vortex line.

Figure 3 shows the dynamics of toroidal vortices propagating in an anomalously dispersive medium ($\beta_2 = -k_0^{-1}$). For the toroidal vortex without longitudinal OAM [Fig. 3(a)], its poloidal energy flow shows a spiral pattern, and thus, the entire morphology is preserved during propagation. However, the introduction of longitudinal OAM disturbs this stable regime. As shown in Figs. 3(b) and 3(c), the vortex line stretches and eventually vanishes, leading to a complete splitting of the toroidal vortex into two separate components. Notably, the direction of the breakup is reversed compared to that observed in the normally dispersive case [Fig. 2].

We now examine the role of longitudinal OAM on the dynamics of toroidal vortices in vacuum ($\beta_2 = 0$). Figure 4(a) shows the toroidal vortex without longitudinal OAM in the process of collapsing, similar to the cases of Fig. 2(a). As the poloidal energy flow is confined to the radial direction, propagation of the toroidal vortex without longitudinal OAM in vacuum over a Rayleigh length will introduce significant distortions. In comparison with Fig. 4(a), the results in Fig. 4(b) indicate that the presence of longitudinal OAM further accelerates the structural distortion. Even more interestingly in Fig. 4(c), as propagation proceeds beyond a Rayleigh length, the vortex line gradually reconstructs, and the entire wavepacket reforms into a well-defined toroidal structure. The reconstructed toroidal vortex contains a polarity reversed vortex line and maintains its morphology over extended propagation in vacuum (see Supplementary Video 1).

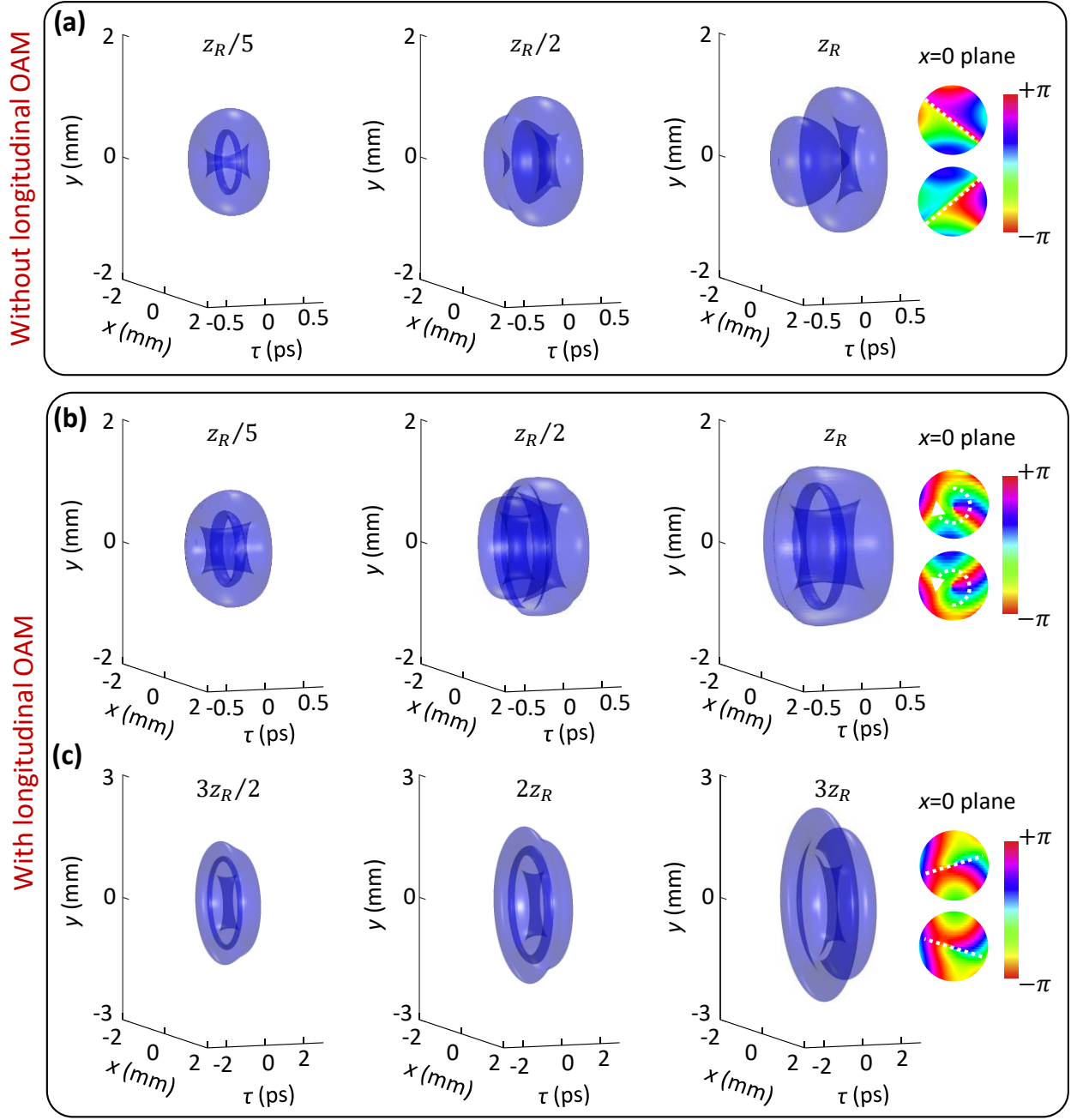

**FIG. 2** Iso-intensity surfaces (10% of the peak intensity) of toroidal vortex wavepacket with $\ell_1 = 1$, (a) $\ell_2 = 0$ and (b) (c) $\ell_2 = 10$, propagating in a normally dispersive medium with $\beta_2 = k_0^{-1}$. Other parameters are the same as in Fig. 1.

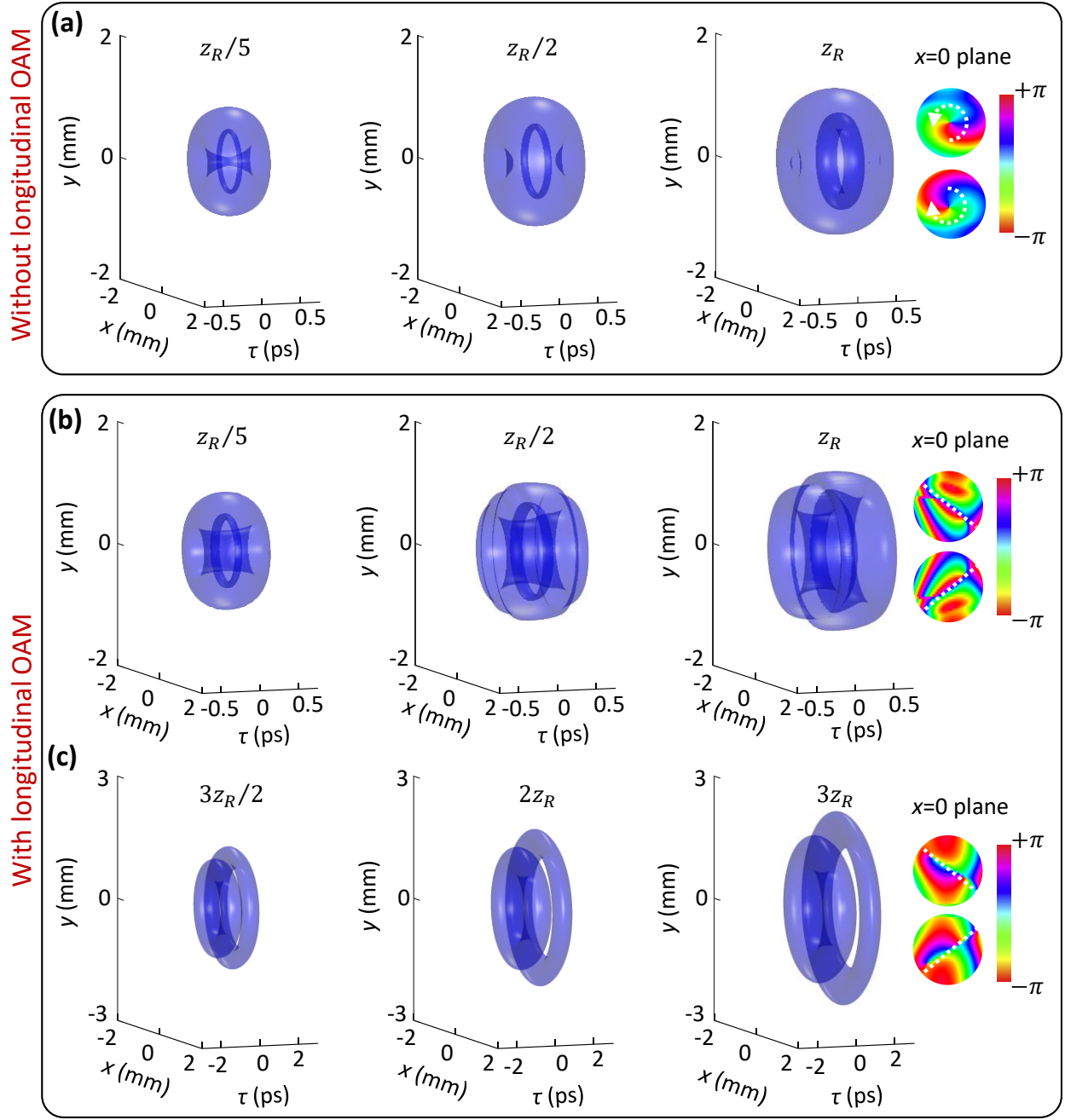

**FIG. 3** Iso-intensity surfaces (10% of the peak intensity) of toroidal vortex wavepacket with $\ell_1 = 1$, (a) $\ell_2 = 0$ and (b) (c) $\ell_2 = 10$, propagating in an anomalously dispersive medium with $\beta_2 = -k_0^{-1}$. Other parameters are the same as in Fig. 1.

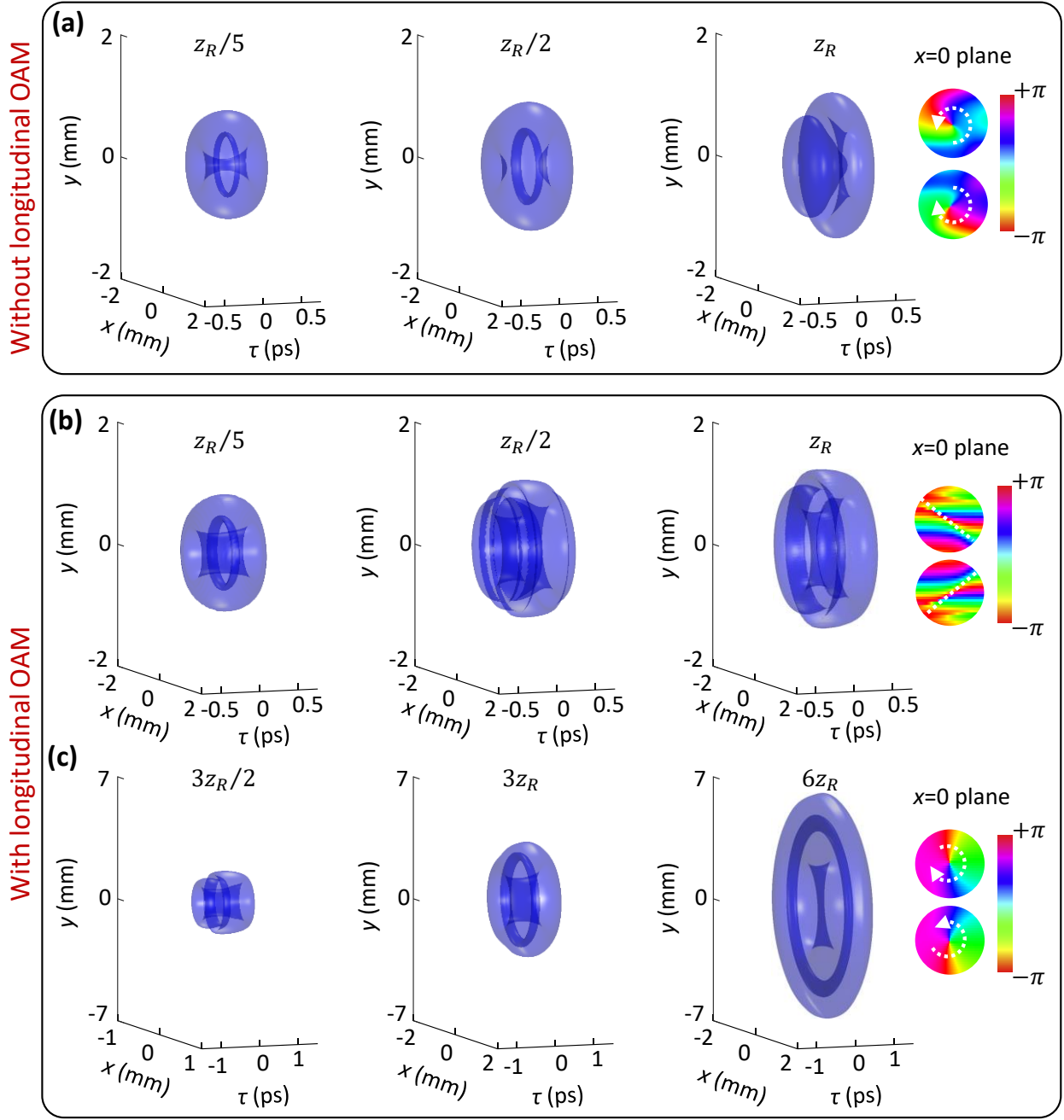

FIG. 4 Iso-intensity surfaces (10% of the peak intensity) of toroidal vortex wavepacket with $\ell_1 = 1$, (a) $\ell_2 = 0$ and (b) (c) $\ell_2 = 10$, propagating in vacuum with $\beta_2 = 0$. Other parameters are the same as in Fig. 1.

## 3. Experimental verification and analysis

The experimental realization of a photonic toroidal vortex can proceed by bending an STOV tube in a conformal mapping system [21,30]. We conducted an experiment that combines 2D pulse shaping with conformal mapping as illustrated in Fig. 5, to verify the simulation results. First, an input pulse centered at 1012nm with a ~20nm bandwidth (Appendix D) is incident onto a 2D pulse shaper. The pulse shaper consists of a grating (1,200 lines/mm), a cylindrical lens with a 10cm focal length along the $x$ axis, and a helical phase pattern (Figure 5 left, in which a group delay dispersion (GDD) phase is added). The modulated pulsed beam is then reflected back, and the grating reconstructs the pulse, generating a near-field STOV pulse with a spatial width of ~250μm. The STOV pulse then travels through an afocal cylindrical beam expander (the ratio of focal length is 2cm:14cm) and stretches in the direction of the vortex line, forming an STOV tube along the $x$-axis. Second, we perform Cartesian to log-polar coordinate transformation to this STOV tube, using dual phase holograms according to conformal mapping theory. In the second stage, the second SLM 2 transforms the STOV tube into a ring-shaped toroidal vortex after propagating $d$=23cm. This is followed by another phase hologram that further shapes the field and applies an additional helical phase of $\ell_2 = 10$, collimating it and completing the afocal transformation. This process ultimately forms a swirl photonic toroidal vortex immediately after the third SLM 3 (at $z$=0). The propagated toroidal vortex at $z$=0.8m is characterized by off-axis interference with a dechirped reference pulse (~100fs). We propagate the generated toroidal vortex in "virtual" dispersive media, where time diffraction affects the pulse similarly to how dispersion influences optical pulses [31,32]. Consequently, the dispersion effect is introduced in the time-frequency domain by modifying the GDD phase in the SLM 1. Finally, the 3D complex field of the wavepacket can be reconstructed from a series of time-delayed interference fringes (Appendix D).

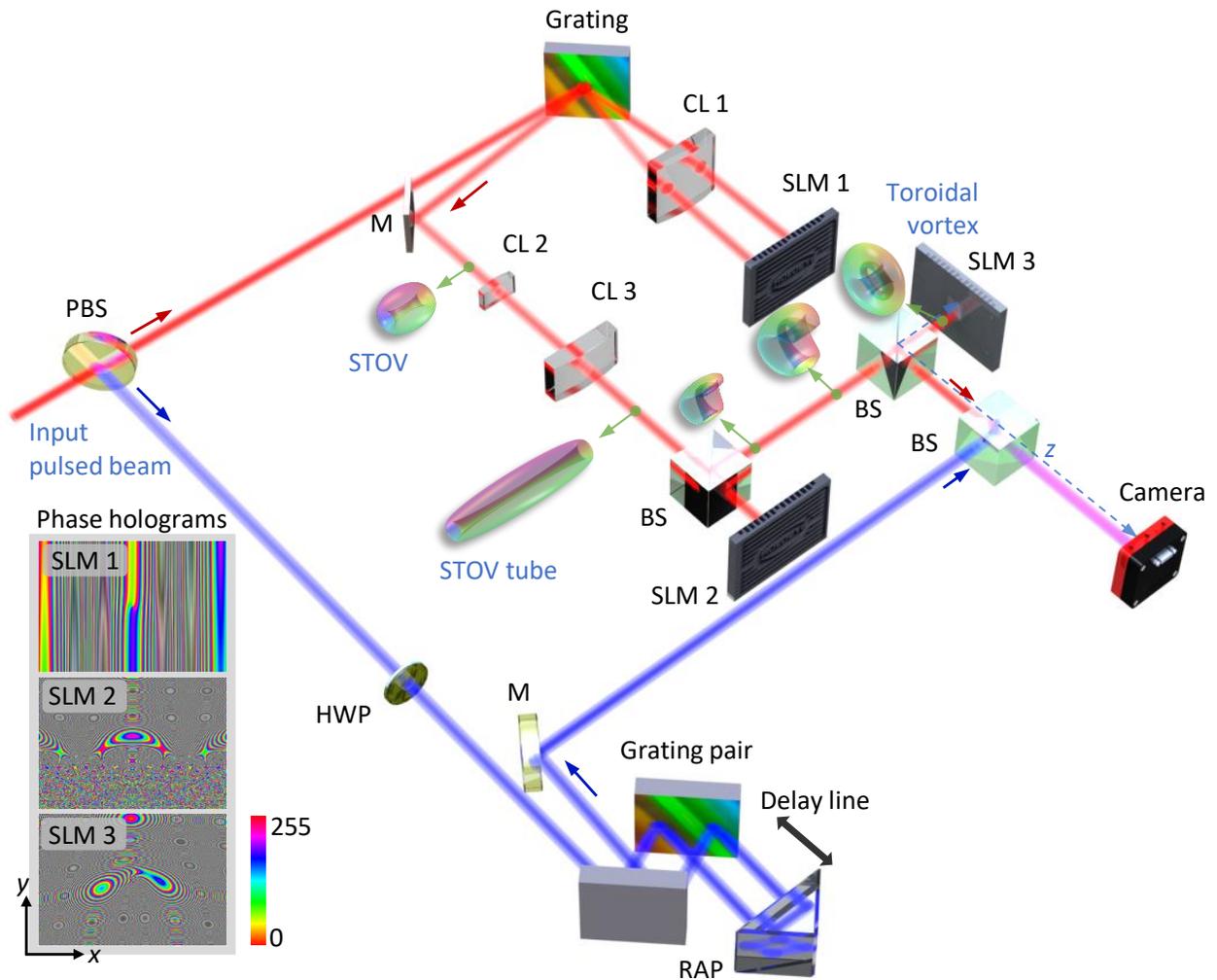

**FIG. 5 Schematic of the experimental setup to synthesize and propagate photonic toroidal vortices. PBS: polarization beam splitter, HWP: half-wave plate, CL 1-3: cylindrical lens, SLM 1-3: spatial light modulator, BS: beam splitter, M: mirror, RAP: right-angle prism. The phase pattern imparted by SLM 1 and SLM 3 includes an additional group delay dispersion phase and a helical phase, respectively.**

Figure 6(a) shows the experimental results of the synthesized toroidal vortex wavepacket of $\ell_1 = 1$ at the source ($z=0$ reflected from the SLM 3; the toroidal radius equals 1mm). For better visualization, semitransparency is applied to the outer torus so that the inside ring-shaped vortex line is clearly shown. The vortex line is colored in red to make the isolated vortex line stand out.

The spatial radius in the toroidal plane of the toroidal vortex is 1mm and 250μm in the poloidal plane, corresponding to the Rayleigh length $z_R$=245mm. Direct propagation of such a toroidal vortex wavepacket in air over long distances results in severe distortions. As shown in Fig. 6(b), the wavepacket undergoes splitting and collapse, leading to the loss of its toroidal structure. For a toroidal vortex carrying longitudinal OAM of $\ell_2 = 10$, as seen in Fig. 6(c), a well-defined toroidal structure is reconstructed and the isolated vortex line reappears. The effect of material dispersion on the wavepackets is "virtually" implemented by adjusting the GDD phase on the pulse shaper. Figures 6(d) and 6(e) show the toroidal vortices with $\ell_2 = 10$ under different dispersion cases. The change in GDD value corresponds to a propagation distance of 0.8m in a dispersive medium of $|\beta_2|$ =160fs$^2$/mm (approximately $|\beta_2| = k_0^{-1}$). In these cases, the wavepackets split reversely in time, and the vortex lines disappear. The renascent toroidal vortex in free space is distinguished by torus-colors different from those of the dispersion-associated breakdown structures. All experimental results are consistent with the simulations presented in Figs. 2-4.

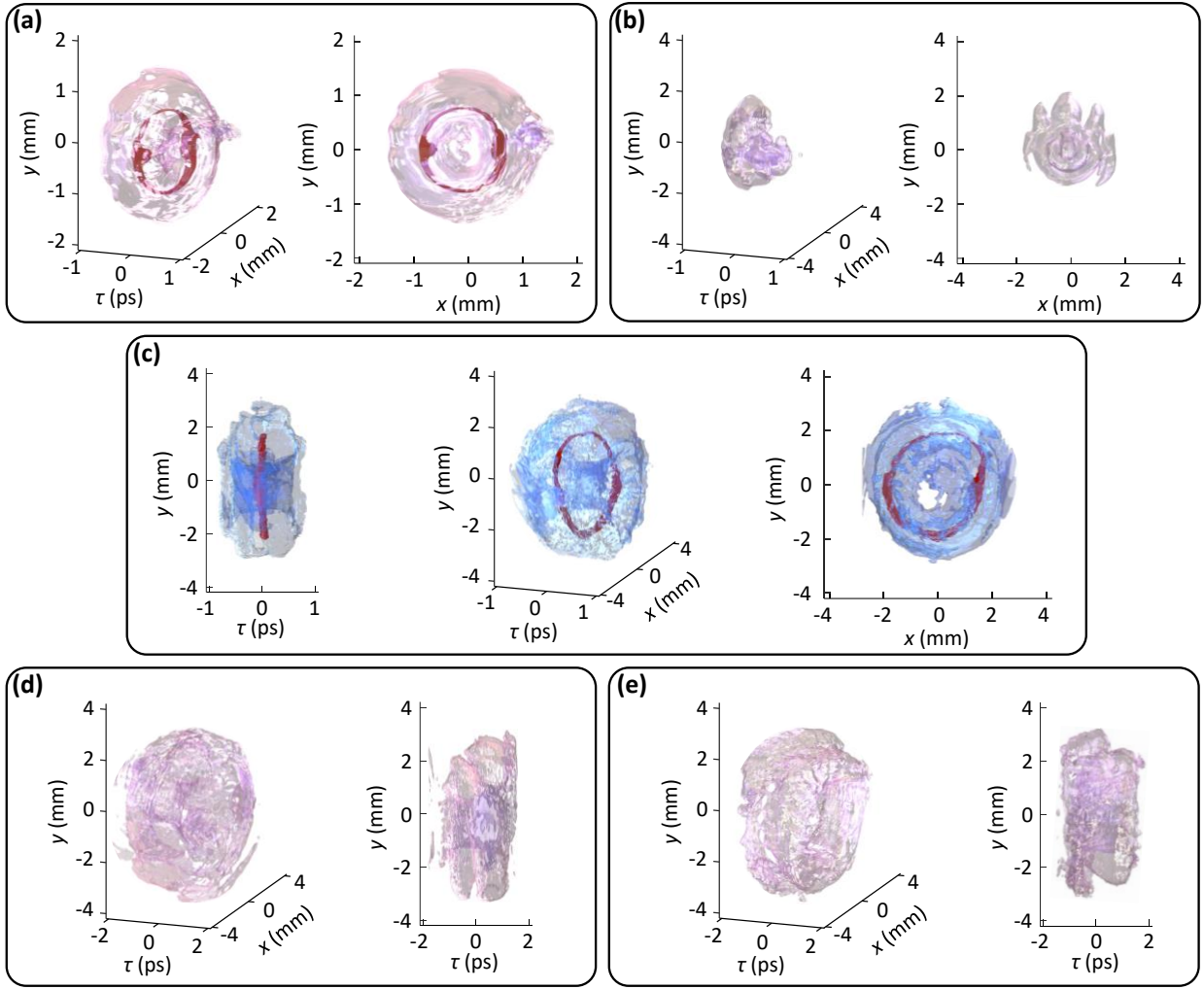

**FIG. 6 Experimental observations of dispersive dynamics of photonic toroidal vortices. (a) 3D iso-intensity profile in different views of the synthesized toroidal vortex source with $\ell_1 = 1$. (b) Iso-intensity profile in different views of a toroidal vortex spreading in vacuum without longitudinal OAM. (c) 3D iso-intensity profile in different views of a renascent toroidal vortex with $\ell_2 = 10$ spreading in vacuum. (d) Iso-intensity profile in different views of a toroidal vortex with $\ell_2 = 10$ spreading in a normal dispersion with a GDD=+0.15ps$^2$. (e) 3D iso-intensity profile in different views of a toroidal vortex with $\ell_2 = 10$ spreading in an anomalous dispersion with a GDD=-0.15ps$^2$. The iso-values are set to (a) 2.5% and (b)-(e) 1% of the peak intensity. The results in (b)-(e) are measured at 0.8m, approximately $3z_R$.**

## 4. Conclusion and discussions

In summary, we demonstrate the propagation dynamics of a class of spatiotemporal toroidal vortex mediated by both transverse and longitudinal OAM. Without the support of longitudinal OAM, the toroidal vortex follows typical STOV propagation dynamics with its structure rapidly deforms and collapses in free space. Toroidal vortices endowed with longitudinal OAM exhibit propagation instability in both normally and anomalously dispersive media, wherein the disappearance of the vortex line results in temporal splitting and disintegration of the wavepacket. What is more interesting is that a photonic toroidal vortex with swirling flow in vacuum undergoes the vanishing and subsequent reconstruction of its vortex line. The renascent toroidal vortex can then stably propagate over an arbitrarily long distance in free space. Owing to its propagation-robust morphology, the photonic toroidal vortex can serve as an effective energy carrier for realizing directional information transfer. These findings hold potential implications for novel photonic topologies engineering, as well as for the development of robust optical encoding, free-space information transmission, remote sensing and ultrafast light-matter interactions.

Finally and notably, some of the phenomena described for the propagation dynamics of toroidal vortices of light have been observed also in fluidics. A toroidal vortex evolved in an anomalously dispersive medium exhibits a spiral energy flow in the poloidal plane—resembling that of a vortex ring in an inviscid incompressible fluid—demonstrating a structurally sustained evolution during propagation [33]. The energy flow scenario of a toroidal vortex with longitudinal OAM is similar to that observed for a swirling vortex ring in a rotating fluid. The swirl flow in the toroidal plane reduces the poloidal wavy deformation of the vortex ring and preserves its ring structure. Consequently, the collapse of the ring structure due to the poloidal wavy deformation as seen in vortex rings without swirl does not appear [34]. Then the traveling distance of a vortex ring can be extended using the swirl flow under certain conditions. While the presence of a swirl component of velocity makes the edge of the ring roll up, splitting and peeling off, thus preventing the vortex line from forming [35-37]. This paper also considers distinct forms of energy flow associated with toroidal vortices, characterized as "saddle"-like in normal dispersion and "parallel" in vacuum. Although such structures are difficult to realize in fluidic systems, they are of considerable interest in optics owing to their distinctive propagation dynamics and field configurations, including phenomena such as vortex line reconstruction and robust wavepacket transport.

# APPENDICES

## APPENDIX A: ENERGY DENSITY AND FLOW OF AN STOV PULSE WITH DISTINCT GROUP VELOCITY DISPERSION

An STOV pulse, whose spatiotemporal envelope at the source ($z=0$) is described by a spatiotemporal Laguerre-Gaussian mode, is given by

$$\Psi(x,z,\tau) = \left(\frac{\sqrt{2}\rho}{w_0}\right)^{|\ell|} \exp\left(-\frac{\rho^2}{w_0^2}\right) L_p^{|\ell|}\left(\frac{2\rho^2}{w_0^2}\right) \exp(-i\ell\varphi), \tag{A1}$$

where $\rho = \sqrt{\tau^2/w_\tau^2 + x^2/w_x^2}$, $\varphi = \tan^{-1}(w_\tau x/w_x \tau)$ and $w_\tau/w_x$ quantifies the ratio of the width of time to space. $w_0$ characterizes the beam width. $L_p^{|\ell|}(\cdot)$ denotes associated Laguerre polynomials of the order $p + |\ell|$. Figures 7(a) and 7(b) display the energy density and phase distributions in the space-time plane of a typical STOV with $p=0$ and $\ell=+1$ at $z=0$. As described by Eq. (C8) and Eq. (C9) in Appendix C, the energy flow is jointly determined by diffraction and dispersion effects. Figures 7 (c)-(e) show the calculated energy density flow $J_{\tau-x}$ in the space-time plane, obtained from Eq. (C7) with various $\beta_2$. It can be observed that the energy density flow of an STOV pulse with $\beta_2 = k_0^{-1}$ exhibits a "saddle" pattern with respect to the singularity. In this case, the STOV cannot sustain intrinsic transverse OAM and undergoes diagonally splitting during propagation because of a strong spatiotemporal astigmatism, as depicted in Figs. 7(f1)-(f3). Beyond a propagation distance approaching the Rayleigh length, the STOV wavepacket exhibits significant degradation, evolving into a multi-lobed structure as illustrated in Fig. 7(f3). But for an STOV propagates in vacuum with $\beta_2 = 0$, the energy flow is restricted to $x$-axis [see Fig. 7(d)]. In this scenario, the STOV primarily expands along the $x$-axis, evolving into a structure characterized by multiple dark regions [Fig. 7(g1)-(g3)], exhibiting an overall shape of a tilted Hermite-Gaussian mode[38]. When $\beta_2 = -k_0^{-1}$, the spatial diffraction and temporal dispersion are balanced, the resulting energy flow of an STOV pulse exhibits a "spiral" pattern that circulates around the singularity. This energy flow enables the STOV pulse to retain a well-defined, donut-shaped intensity profile and a preserved helicoid phase structure over a distance [see Figs. 7(h1)-(h3)]. As a summary, the cross-section structure of a spatiotemporal wavepacket is directly governed by their energy flow dynamics, which are influenced by the interplay of dispersion and diffraction.

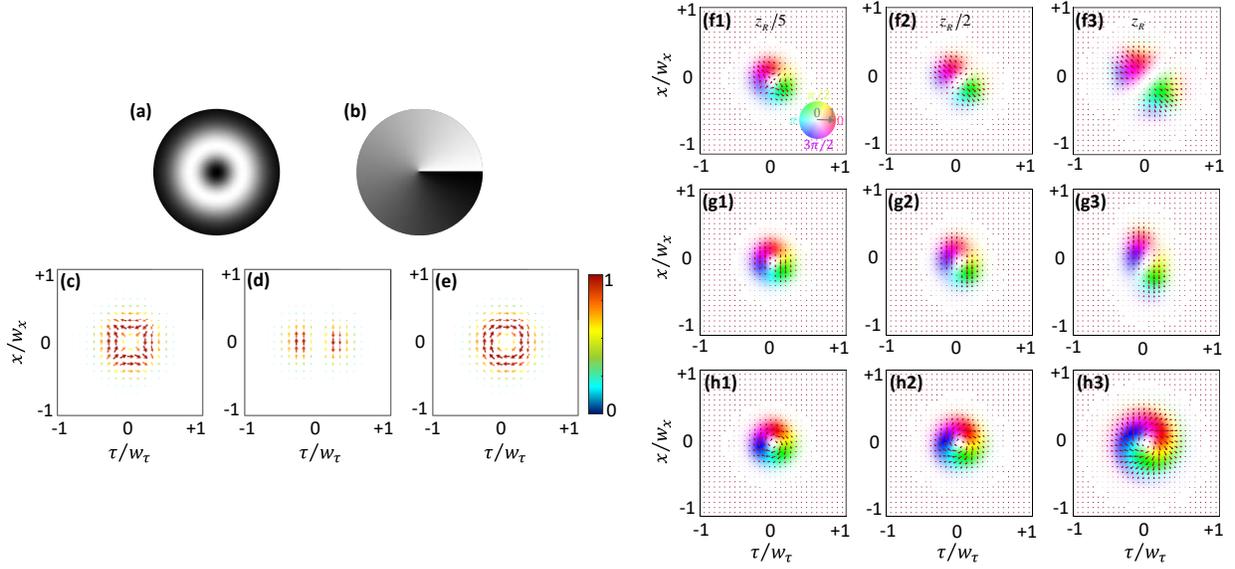

FIG. 7 Energy density (a) and phase (b) distribution of an STOV ($\ell=+1$) wavepacket (defined in the space-time plane). (c)-(e) Calculated energy density flow $\mathbf{J}_{\tau-x}$ of an STOV wavepacket with (c) $\beta_2 = k_0^{-1}$, (d) $\beta_2 = 0$ and (e) $\beta_2 = -k_0^{-1}$. (f)-(g) Complex field of an STOV ($\ell=+1$) wavepacket at different propagation distances with (f1)-(f3): $\beta_2 = k_0^{-1}$; (g1)-(g3): $\beta_2 = 0$ and (h1)-(h3) $\beta_2 = -k_0^{-1}$. The propagation distances from left to right are $0.2z_R, 0.5z_R$ and $z_R$. In each plot, the saturation and hue denote energy density and phase distribution respectively. The energy flow in the local time frame is indicated by the red arrows. The parameters are $\lambda_0 = 800$nm $w_\tau =1$ps, $w_x = 1$mm, $w_0 =0.25$.

**APPENDIX B: ENERGY FLOW OF TOROIDAL VORTICES**

Figure 8 illustrates the calculated energy flow of toroidal wavepackets of $\ell_2 = 0$ with various

$\beta_2$. In Fig. 8(a), a toroidal vortex lacking longitudinal OAM exhibits an energy flow in the poloidal plane identical to that of an STOV pulse. However, in Fig. 8(b), the toroidal vortex with longitudinal OAM, possessing both transverse [$\langle \hat{L}_x \rangle$ and $\langle \hat{L}_y \rangle$ in the poloidal plane] and longitudinal [$\langle \hat{L}_\tau \rangle \propto \ell_2 \neq 0$ in the toroidal plane] OAM, displays an energy density flow with a distinct toroidal component. We would like to mention that, as described in the main text, the energy flow patterns of photonic toroidal vortices with or without longitudinal OAM in anomalous dispersion closely resemble those of vortex rings with or without swirl in fluid dynamics, exhibiting analogous motion and dynamic behavior[33-37].

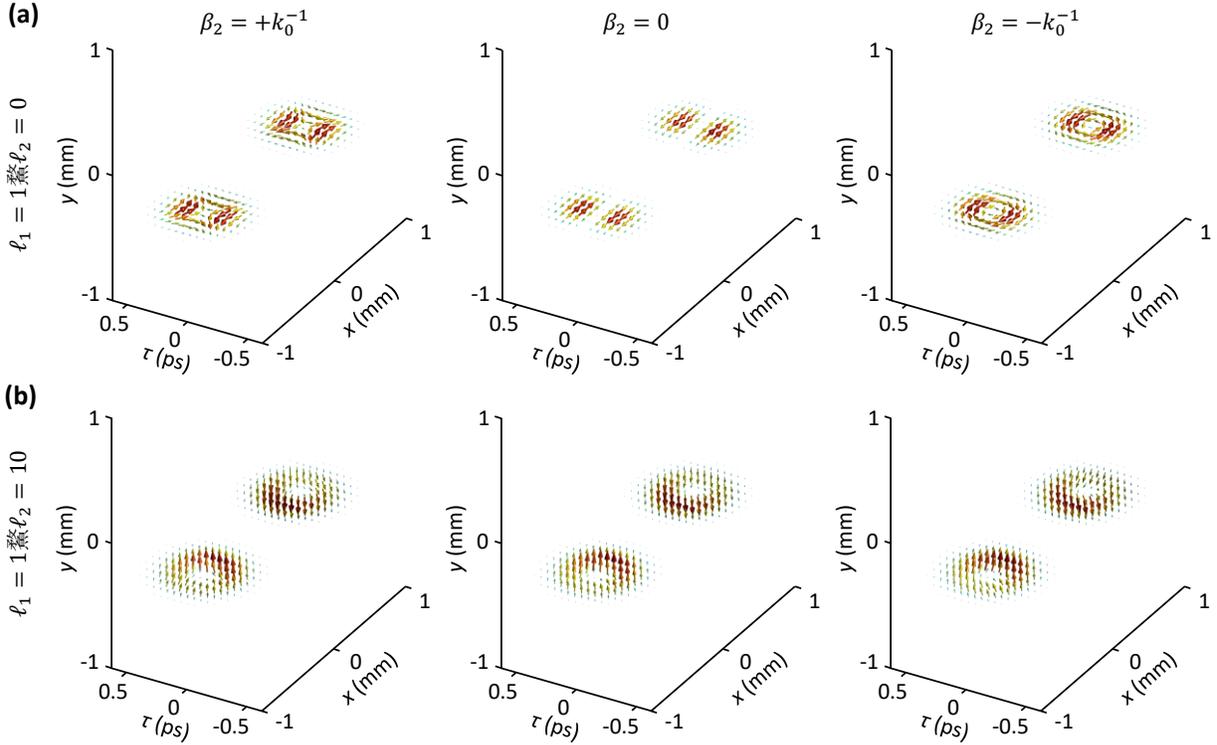

FIG. 8 Calculated energy density flow [Eq. (C7)] of a photonic toroidal vortex (a) without or (b) with longitudinal OAM under various $\beta_2$. Take a specific poloidal plane ($y=0$) as an example. Other parameters are the same as in Fig. 1.

**APPENDIX C: NUMERICAL CALCULATION AND MOMENTUM FLOW OF**

## PROPAGATION FOR A FULL SPACE-TIME WAVEPACKET IN DISPERSIVE MEDIA

Based on the electromagnetic theory of light and Maxwell's equations, the propagation of a light pulse, characterized by a scalar complex field $E(x, y, z; t)$ of center carrier frequency $\omega_0$, in an isotropic, transparent, passive linear medium is governed by the following wave equation

$$\left(\frac{\partial^2}{\partial x^2} + \frac{\partial^2}{\partial y^2} + \frac{\partial^2}{\partial z^2}\right) E(x, y, z; t) - \frac{1}{c^2}\frac{\partial^2 E(x, y, z; t)}{\partial t^2} = 0, \quad (C1)$$

where $c$ is the speed of light in vacuum. We assume that the light pulse is travelling at both paraxial approximation with $|\Delta k| \ll k_0$ and narrow bandwidth approximation with $|\Delta \omega| \ll \omega_0$. With these approximations, the froward ($z>0$) electric field can be expressed as the product of a slowly varying envelop and the carrier oscillation at the central frequency, given by $E(x, y, z; t) = \Psi(x, y, z; t) \exp(-i\omega_0 t + ik_0 z)$, where $k_0 = \omega_0/c = 2\pi/\lambda_0$ is a propagation constant. In the temporal frequency domain, the electric field has $\hat{E}(x, y, z; \omega) = \hat{\Psi}(x, y, z; \Omega) \exp(ik_0 z)$ with the detuning angular frequency $\Omega = \omega - \omega_0$, where the top mark denotes time-only Fourier transform: $\hat{f}(\omega) = \int f(t) \exp(-i\omega t) \, dt$. Taking the replacing operators $\partial E/\partial t = -i\omega_0 \hat{E}$ and $\partial^2 E/\partial t^2 = -\omega_0^2 \hat{E}$, one can obtain the paraxial wave equation in the frequency form

$$\left(\frac{\partial^2}{\partial x^2} + \frac{\partial^2}{\partial y^2}\right)\hat{\Psi} + 2ik_0\frac{\partial\hat{\Psi}}{\partial z} + \left[k^2(\omega) - k_0^2\right]\hat{\Psi} = 0. \quad (C2)$$

It should be noted that the derivation of Eq. (C2) from Eq. (C1) relies on two slowly varying envelope approximations (SVEAs): (i) spatial SVEA, the distance light travels during the duration of a pulse is much larger than its wavelength, i.e., $|\partial^2 E/\partial z^2| \ll |k\partial E/\partial z| \ll |k^2 E|$, and (ii) temporal SVEA, the temporal width of the optical envelop is much larger than its carrier period, i.e., $|\partial^2 E/\partial t^2| \ll |\omega \partial E/\partial t| \ll |\omega^2 E|$. As a result, the second-order derivative $\partial^2 E/\partial z^2$ is negligible with respect to $k\partial E/\partial z$ during the above derivation. In Eq. (C2), the paraxial approximation allows $k^2(\omega) - k_0^2 \approx 2k_0[k(\omega) - k_0]$ and the Taylor expands of $k(\omega)$ reads $k(\omega) = \sum_{m=0}^{+\infty} k^{(m)}(\omega_0)\Omega^m/m!$, where $v_g = 1/k^{(1)}(\omega_0)$ and $\beta_2 = k^{(2)}(\omega_0)$ is group velocity and group velocity dispersion (GVD) at $\omega_0$, respectively. By expanding the first three terms in a Taylor series and applying the inverse operators $-i\Omega\tilde{\Psi} = \partial\Psi/\partial t$, $-\Omega^2\tilde{\Psi} = \partial^2\Psi/\partial t^2$, we can readily obtain

$$\frac{\partial \Psi}{\partial z} = \frac{i}{2k_0}\left(\frac{\partial^2}{\partial x^2} + \frac{\partial^2}{\partial y^2}\right)\Psi - i\frac{\beta_2}{2}\frac{\partial^2 \Psi}{\partial \tau^2}, \tag{C3}$$

where $\tau = t - z/v_g$ is the local variables. A solution to Eq. (C3) is given by:

$$\Psi(x,y,z,\tau) = \frac{\exp\left[ik_0 z + ik_0(x^2+y^2)/2z\right]}{2\pi} \times \iiint \Psi(x_0,y_0,0,\Omega) \\ \exp\left(ik_0\frac{x_0^2+y_0^2}{2z} + i\beta_2\frac{\Omega^2}{2}z\right)\exp\left(-ik_0\frac{xx_0+yy_0}{z} - i\Omega\tau\right)d\Omega dx_0 dy_0, \tag{C4}$$

and an angular spectrum relation

$$\Psi(x,y,z,\tau) = \frac{\exp(ik_0 z)}{8\pi^3} \times \\ \iiint \tilde{\Psi}(k_x,k_y,\Omega)H(k_x,k_y,\Omega)\exp(-ik_x x - ik_y y - i\Omega\tau)d\Omega dk_x dk_y, \tag{C5}$$

where $H(k_x,k_y,\Omega) = \exp\left[-i\left(k_x^2+k_y^2\right)z/2k_0 + i\beta_2\Omega^2 z/2\right]$. For a full space-time optical wavepacket with entangled 3D variables, obtaining closed-form solutions for Eqs. (C4) and (C5) is challenging. The above integral equations are numerically obtainable using the Fast Fourier Transform algorithm. For a scalar wavepacket $\Psi(x,y,\varsigma;\tau)$, the square of its modulus $|\Psi|^2$ represents the energy density at a given position. Integrating over a volume $V$ in the $x$-$y$-$\tau$ space $\int_V |\Psi|^2 dx dy d\tau$, gives the total energy carried by the wavepacket. The total energy satisfies an energy conservation equation, which can be derived by multiplying Eq. (C3) by the complex conjugate of the wavepacket. This equation is expressed as:

$$\frac{\partial |\Psi|^2}{\partial \varsigma} = -\frac{i}{2k_0}\left[\frac{\partial}{\partial x}\left(\Psi\frac{\partial \Psi^*}{\partial x} - \Psi^*\frac{\partial \Psi}{\partial x}\right) + \frac{\partial}{\partial y}\left(\Psi\frac{\partial \Psi^*}{\partial y} - \Psi^*\frac{\partial \Psi}{\partial y}\right)\right]$$
$$-i\frac{\beta_2}{2}\frac{\partial}{\partial \tau}\left(\Psi^*\frac{\partial \Psi}{\partial \tau} - \Psi\frac{\partial \Psi^*}{\partial \tau}\right). \quad (C6)$$

The right-hand side of Eq. (C6) takes an analogous form to the definition of the divergence operator in the $x$-$y$-$\tau$ space[39], as follows

$$\frac{\partial |\Psi|^2}{\partial \varsigma} = -\mathrm{div}\mathbf{J} = -\nabla_\perp \cdot \mathbf{J}_\perp - \frac{\partial \mathbf{J}_\tau}{\partial \tau}, \quad (C7)$$

where $\nabla_\perp = \partial/\partial x\, \vec{x} + \partial/\partial y\, \vec{y}$ denotes the transverse curl operator. Equation (C7) enables the identification of the total energy density flux $\mathbf{J} = \mathbf{J}_\perp + \mathbf{J}_\tau$ through the whole surface of an enclosing volume $V$, is given by $\mathbf{J}_\perp = \frac{i}{2k_0}(\Psi^*\nabla_\perp\Psi - \Psi\nabla_\perp\Psi^*)$ and $\mathbf{J}_\tau = i\frac{\beta_2}{2}\left(\Psi^*\frac{\partial \Psi}{\partial \tau} - \Psi\frac{\partial \Psi^*}{\partial \tau}\right)\vec{\tau}$.

The cross product of the energy density flow $\mathbf{J}$ with position vector $\vec{r} = x\vec{x} + y\vec{y} + \tau\vec{\tau}$ gives an OAM operator below

$$\hat{L}_x = -\frac{i}{k_0}\left(-yk_0\beta_2 \frac{\partial}{\partial \tau} - \tau\frac{\partial}{\partial y}\right),$$
$$\hat{L}_y = -\frac{i}{k_0}\left(\tau\frac{\partial}{\partial x} + xk_0\beta_2 \frac{\partial}{\partial \tau}\right), \quad (C8)$$
$$\hat{L}_\tau = -\frac{i}{k_0}\left(x\frac{\partial}{\partial y} - y\frac{\partial}{\partial x}\right).$$

Equation (C8) can also be expressed in the cylindrical coordinate as follows:

$$\hat{L}_x = -\frac{i}{k_0}\left[-(1+k_0\beta_2)\frac{\partial}{\partial r}r\cos\theta\sin\theta + (k_0\beta_2\sin^2\theta - \cos^2\theta)\frac{\partial}{\partial \theta}\right],$$
$$\hat{L}_y = -\frac{i}{k_0}\left[(1+k_0\beta_2)\frac{\partial}{\partial \rho}\rho\sin\varphi\cos\varphi + (\cos^2\varphi - k_0\beta_2\sin^2\varphi)\frac{\partial}{\partial \varphi}\right], \quad (C9)$$
$$\hat{L}_\tau = -\frac{i}{k_0}\frac{\partial}{\partial \gamma}\vec{\tau},$$

where $r = \sqrt{\tau^2 + y^2}, \theta = \tan^{-1}(y/\tau), \rho = \sqrt{\tau^2 + x^2}, \varphi = \tan^{-1}(x/\tau)$ and $\gamma = \tan^{-1}(y/x)$.

Equation (C9) reveals that the first term of each expression represents extrinsic transverse OAM, while the second term represents intrinsic transverse OAM per photon, determined by not only the phase gradient but also the GVD coefficient. Equation (C9) describes the longitudinal OAM $\hat{L}_\tau$, which is orthogonal to the transverse OAM characterized by $\hat{L}_x$ and $\hat{L}_y$.

**APPENDIX D: INPUT PULSE AND TOROIDAL VORTEX CHARACTERIZATION**

In the experiment, the ultrafast laser is a dispersion-managed mode-locked fiber laser. Figure 9 (a) shows the spectrum of the laser center at 1012nm with a bandwidth of ~20nm. The duration of the generated photonic toroidal vortices, exceeding 1 picosecond, is longer than that of the reference pulse. The complete space-time structure can be reconstructed from a series of time-delayed interference fringes, produced via off-axis (~0.25 degree) interference with a transform-limited reference pulse, as shown in Figures 9(b) and 9(c).

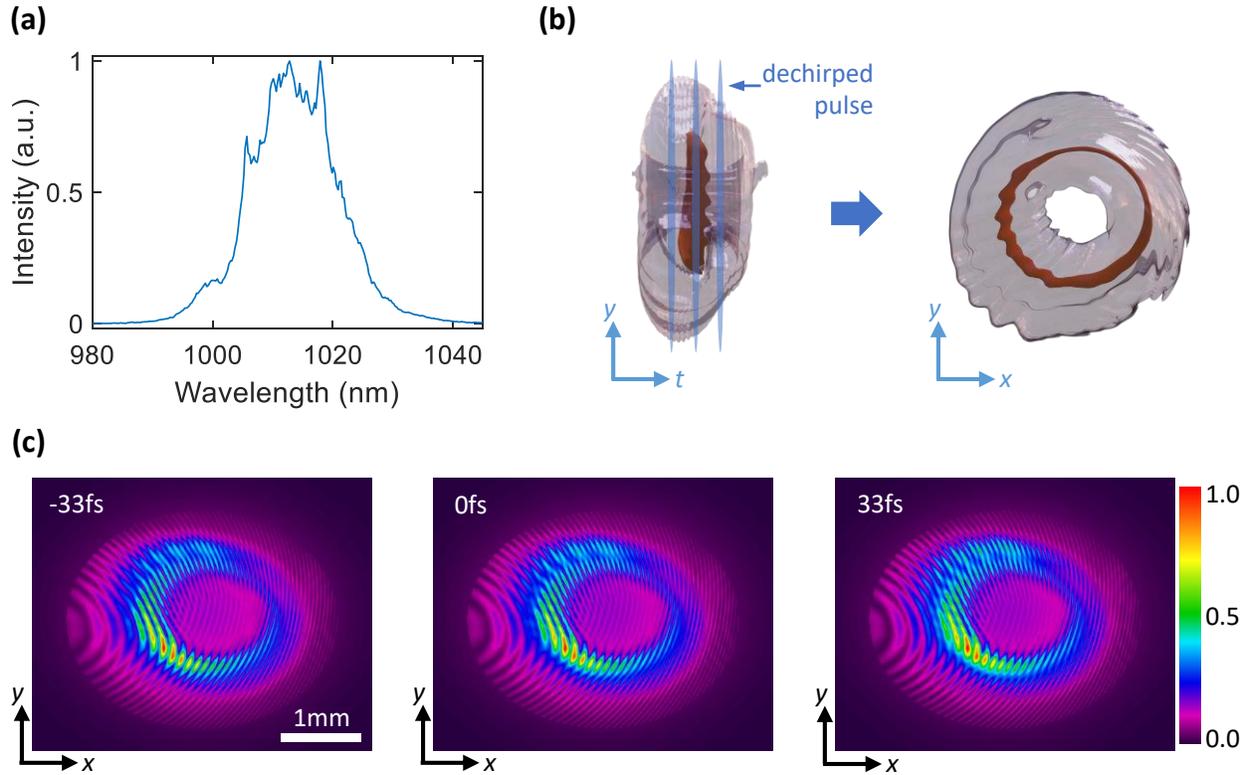

FIG. 9 (a) Spectrum of the fiber laser. (b) Characterization of spatiotemporal toroidal vortex. (c)

Some representative interference patterns of the generated toroidal vortex of $\ell_1 = 1$ and $\ell_2 = 10$ with the dechirped pulsed beam at different times.

**Funding.** This work was supported by the National Key Research and Development Program of China (2022YFA1404800 [Y.C.]), the National Natural Science Foundation of China (12434012 [Q.Z.], 12192254 [Y.C.], 92250304 [Y.C.], W2441005 [Y.C.], 12374311 [C.L.], 12474336 [Q.C.]), the Natural Science Foundation of Shandong Province (ZR2024QA216 [J.L.], ZR2023YQ006 [C.L.]), the Shanghai Science and Technology Committee (24JD1402600 [Q.Z.], 24QA2705800 [Q.C.]), and the Taishan Scholars Program of Shandong Province (tsqn202312163 [C.L.]).

**Disclosures.** The authors declare no competing interests.

**Data availability.** The data that support the findings of this study are available from the authors upon reasonable request.

**Author contributions.** X. Liu conceived the project, designed the experiments, and performed the simulations. X. Liu and N. Zhang conducted the experiments. X. Liu, Q. Cao, J. Liu, and C. Liang both analyzed the data. X. Liu, Q. Zhan, and Y. Cai drafted and finalized the manuscript. Q. Zhan and Y. Cai supervised the project. All authors contributed to the discussion of the manuscript.